\documentstyle[12pt]{article}

\begin{document}
\newcommand{\nd}{\noindent}
\newcommand{\beq}{\begin{equation}}
\newcommand{\eeq}{\end{equation}}
\newcommand{\barr}{\begin{eqnarray}}
\newcommand{\earr}{\end{eqnarray}}
\newcommand{\NP}[1]{{\it Nucl.\ Phys.}\ {\bf #1}}
\newcommand{\PL}[1]{{\it Phys.\ Lett.}\ {\bf #1}}
\newcommand{\PR}[1]{{\it Phys.\ Rev.}\ {\bf #1}}
\newcommand{\PRL}[1]{{\it Phys.\ Rev.\ Lett.}\ {\bf #1}}
\newcommand{\MPL}[1]{{\it Mod.\ Phys.\ Lett.}\ {\bf #1}}
\newcommand{\SNP}[1]{{\it Sov.\ J.\ Nucl.\ Phys.}\ {\bf #1}}

\begin{flushright}
DFTT47/94 \\
INFNCA-TH-94-21 \\
October 1994
\end{flushright}

\vspace{0.2in}

\renewcommand{\thefootnote}{\fnsymbol{footnote}}

\begin{center}
{\LARGE Single Spin Asymmetry \\
In Inclusive Pion Production%
\footnote{Talk presented by F. Murgia at the XI International
Symposium on High Energy Spin Physics, September 15-22,
Bloomington, Indiana.}}
\end{center}

\vskip12pt

\begin{center}
{\large M. Anselmino$^\dagger$, M.E. Boglione$^\dagger$
and F. Murgia$^\ddagger$ }
\end{center}

\vskip12pt

\begin{center}
{\small\it $^\dagger$Dipartimento di Fisica Teorica, Universit\`a
di Torino and \\
Istituto Nazionale di Fisica Nucleare, Sezione di Torino, \\
Via P. Giuria 1, I-10125 Torino, Italy \\
$^\ddagger$Istituto Nazionale di Fisica Nucleare,
Sezione di Cagliari, \\
Via Ada Negri 18, I-09127 Cagliari, Italy }
\end{center}

\vskip30pt

\begin{center}
\nd \parbox{5.0in}{\small {\bf Abstract.}
It is shown how the single spin asymmetry observed in inclusive pion
production is related, in the helicity basis, to the imaginary part
of the product of two different distribution amplitudes, rather than
to the usual quark and gluon distribution functions; there is then no
reason why it should be zero even in massless perturbative QCD, provided
the quark intrinsic motion is taken into account. A simple model is
constructed which reproduces the main features of the data.}
\end{center}

\vskip12pt

Spin physics in large $p_T$ inclusive hadronic processes has
unique features; not only it probes the internal structure of hadrons,
but, as spin dependent observables involve delicate interference
effects among different amplitudes, it tests the theory
at a much deeper level than unpolarized processes.

We consider here the single spin asymmetry in inclusive pion production
in $p-p$ collisions, $p^{\uparrow} + p \to \pi+X$. Let the
two protons move along the $\hat z$-axis in their c.m. frame and
$\hat x$-$\hat z$ be the scattering plane. The proton moving in the
$+\hat z$ direction is polarized transversely to the scattering plane,
{\it i.e.} along ($\uparrow$) or opposite ($\downarrow$) the $\hat y$-axis.
The single spin asymmetry $A_N$ is then defined by:

\beq
A_N(x_F,p_T) = \frac{d\sigma^\uparrow - d\sigma^\downarrow}
                    {d\sigma^\uparrow + d\sigma^\downarrow}
\label{andef}
\eeq
\nd
where $d\sigma$ is the differential cross section and $\uparrow$,
$\downarrow$ refer to the proton spin directions; we denote by $p_L$ and
$p_T$ the c.m. longitudinal and transverse pion momentum respectively;
$x_F = 2 p_L/\sqrt{s}$ is the Feynman variable and $\sqrt{s}/2$ is the
c.m. energy of each incident proton.

Several experimental results are available on $A_N$ (for a list of
references see \cite{abm}); the E704 Collaboration has produced the most
recent, high energy ones ($\sqrt{s}/2 \simeq 10$ GeV).
Two sets of measurements are relevant to our analysis:

i)  $A_N(x_F,p_T)$ for $p^\uparrow + p \to \pi^\pm,\pi^0 + X$,
    {\it vs.} $x_F$ in the $p_T$ range $0.7 \leq p_T \leq 2.0$
    GeV/$c$ \cite{e704a,e704b}; these data show intriguing $x_F$
    dependence (Fig. 1).

ii) $A_N(x_F,p_T)$ for $p^\uparrow + p \to \pi^0 + X$, as a function
    of $p_T$ (up to $p_T \simeq 4$ GeV/$c$), in the central region ($|x_F|
    \leq 0.1$) \cite{nur}; in this case no $p_T$ dependence seems to be
    observed and $A_N \simeq 0$ in the whole $p_T$ range (notice that
    this updates and corrects some previous results of the same
    collaboration \cite{e704c}).

A naive generalization of the QCD-factorization theorem suggests that
the single spin asymmetry can be written qualitatively as:

\beq
A_N  \sim \sum_{ab \to cd} \Delta_T G_{a/p} \otimes G_{b/p}
\otimes \hat a_N \hat \sigma_{ab\to cd} \otimes D_{\pi/c}
\label{qual}
\eeq
\nd
where $G_{a/p}$ is the parton distribution function, that is
the number density of partons $a$ inside the proton, and
$\Delta_T G_{a/p} = G_{a^{\uparrow}/p^{\uparrow}} -
G_{a^{\downarrow}/p^{\uparrow}}$ is the difference between the number
density of partons $a$ with spin $\uparrow$ in a proton with spin $\uparrow$
and the number density of partons $a$ with spin $\downarrow$ in a proton
with spin $\uparrow$; $D_{\pi/c}$ is the number density of pions resulting
from the fragmentation of parton $c$; $\hat a_N$ is the single spin
asymmetry relative to the $a^{\uparrow}b \to cd$ elementary process and
$\hat \sigma$ is the cross-section for such process.

\renewcommand{\thefootnote}{\arabic{footnote}}

The usual argument is then that the asymmetry (\ref{qual}) is bound to be
very small because $\hat a_N \sim \alpha_s m_q /\sqrt s$ where $m_q$ is
the quark mass. This originated the widespread opinion that single spin
asymmetries are essentially zero in perturbative QCD.

However, it has become increasingly clear in the last years that such
conclusion need not be true because subtle spin effects might modify Eq.~
(\ref{qual}). Such modifications should take into account the parton
transverse motion, higher twist contributions and
possibly non perturbative effects hidden in the spin dependent distribution
and fragmentation functions. Several models have been proposed which
differ in practice by which part of $A_N$, Eq.~(\ref{qual}),
is responsible for these effects: $\Delta_T G_{a/p}$ \cite{siv,qiu,meng},
$\hat \sigma$ \cite{szwed,efre} or $D_{\pi/c}$ \cite{col,artru}.

We briefly discuss here a reformulation of Eq.~(\ref{qual}) in the helicity
basis, which is more suitable for applying the factorization theorem and
which allows to formulate a model for the spin dependence of the quark
distributions \cite{abm}. Our approach is reminiscent of that of
Ref.~\cite{siv}.

In the helicity basis the differential cross-section for the inclusive
process $p_1(\lambda_1)+p_2(\lambda_2) \to \pi + X(\lambda_X)$ can be
written in terms of helicity amplitudes as:
\beq
d\sigma \sim \sum_{X,\lambda_X}~\sum_{\lambda_1,\lambda'_1,
\lambda_2,\lambda'_2} M_{\lambda_X;\lambda_1,\lambda_2}
\ \rho_{\lambda_1,\lambda_2;\lambda'_1,\lambda'_2}(p_1,p_2)
\ M^\ast_{\lambda_X;\lambda'_1,\lambda'_2}
\label{ds}
\eeq
\nd
where the sum over $X$ includes also a phase space integral for the undetected
particles and the matrix $\rho$ is the helicity density matrix describing
the polarization state of the initial protons. In our case
$p_2$ is unpolarized, while $p_1$ is transversely polarized along
$\pm \hat y$ direction, so that Eq.~(\ref{andef}) becomes

\beq
A_N = 2~\frac{\sum_{_{X,\lambda_X,\lambda_2}}
\mbox{Im}[M_{\lambda_X;+,\lambda_2}M^\ast_{\lambda_X;-,\lambda_2}]}
{\sum_{_{X,\lambda_X,\lambda_1,\lambda_2}}
|M_{\lambda_X;\lambda_1,\lambda_2}|^2} \cdot
\label{anamp}
\eeq
\nd
Eq.~(\ref{anamp}) shows how a non zero value of $A_N$ implies non zero
interference effects between two amplitudes which only differ by one
helicity index; its denominator, instead, proportional to $d\sigma^{\uparrow}
+ d\sigma^{\downarrow} = 2\,d\sigma^{unp}$, only depends on moduli squared of
amplitudes and can be written in the parton model as:

\beq
d\sigma^{unp} \sim \sum_{abcd} \int dx_a dx_b ~
\frac{1}{x_c} \, G_{a/p}(x_a) \, G_{b/p}(x_b) \,
\frac{d\hat\sigma}{d\hat t}(ab\to cd) \, D_{\pi/c}(x_c)\,.
\label{dsunp}
\eeq

In order to express the numerator of Eq.~(\ref{anamp}) in terms
of parton interactions we have to define
${\cal G}^{a/h}_{\lambda_{X_h},\lambda_a;\lambda_h}(x_a,
\mbox{\boldmath $k$}_{\perp_a})$ as the helicity distribution amplitude
for the process $h(\lambda_h) \to a(\lambda_a) + X_h(\lambda_{X_h})$,
where $\mbox{\boldmath $k$}_{\perp_a}$ is the transverse momentum of
the parton $a$ inside the hadron $h$; these amplitudes are related to
the unpolarized partonic distribution function by:

\beq
G_{a/p}(x_a) = \sum_{X_p,\lambda_{X_p}}
\int d\mbox{\boldmath $k$}_{\perp a}\left\{
|{\cal G}^{a/p}_{\lambda_{X_p},+;+}(x_a,\mbox{\boldmath $k$}_{\perp a})|^2 +
|{\cal G}^{a/p}_{\lambda_{X_p},-;+}(x_a,\mbox{\boldmath $k$}_{\perp a})|^2
\right\} \,.
\label{g}
\eeq

By applying the same steps which lead to the partonic
expression of $d\sigma^{unp}$ we get for the numerator
of $A_N$ an expression similar to Eq.~(\ref{dsunp}), with $G_{a/p}(x_a)$
replaced by $\int d\mbox{\boldmath $k$}_{\perp a} \,
I^{a/p}_{+-}(x_a,\mbox{\boldmath $k$}_{\perp a})$, where

\beq
I^{a/p}_{+-}(x_a,\mbox{\boldmath $k$}_{\perp a}) \equiv
\sum_{X_p,\lambda_{X_p}} \mbox{Im}[
{\cal G}^{a/p}_{\lambda_{X_p},+;+}(x_a,\mbox{\boldmath $k$}_{\perp a}) \,
{\cal G}^{a/p\,*}_{\lambda_{X_p},+;-}(x_a,\mbox{\boldmath $k$}_{\perp a})]\,.
\label{iap}
\eeq

Notice that $I^{a/p}_{+-}(x_a,\mbox{\boldmath $k$}_{\perp a})$ has to vanish
for $\mbox{\boldmath $k$}_{\perp a}=0$, as required by
helicity conservation in the forward direction; moreover, since
$I^{a/p}_{+-}(x_a,\mbox{\boldmath $k$}_{\perp a})$ is an odd function of
$\mbox{\boldmath $k$}_{\perp a}$%
\addtocounter{footnote}{-1}\footnote{This is more easily seen if
we observe that
our $I^{a/p}_{+-}(x_a,\mbox{\boldmath $k$}_{\perp a})$ equals
$\Delta^N G_{a/p^{\uparrow}}(x_a,\mbox{\boldmath $k$}_{\perp a})$=
$\sum_{\lambda_a}\bigl\{G_{a(\lambda_a)/p^{\uparrow}}
(x_a,\mbox{\boldmath $k$}_{\perp a})$ $-
 G_{a(\lambda_a)/p^{\uparrow}}(x_a,-\mbox{\boldmath $k$}_{\perp a})\bigr\}$
defined by Sivers \cite{siv}.},
we must keep into account
$\mbox{\boldmath $k$}_{\perp a}$ effects also in the partonic cross
sections, otherwise we are left with
$\int d\mbox{\boldmath $k$}_{\perp a}~
I^{a/p}_{+-}(x_a,\mbox{\boldmath $k$}_{\perp a})
=0$. Then

\beq
\int \!d\mbox{\boldmath $k$}_{\perp a}
I^{a/p}_{+-}(x_a,\mbox{\boldmath $k$}_{\perp a})
\frac{d\tilde \sigma}{d\tilde t}(\mbox{\boldmath $k$}_{\perp a}) =
\!\int
\!\!\!\!\!\!\!\!\!\!\!
\raisebox{-.5truecm}{$\scriptstyle (k_{\perp a})_x > 0$}
\!\!\!\!\!\!\!\! d\mbox{\boldmath $k$}_{\perp a}
I^{a/p}_{+-}(x_a,\mbox{\boldmath $k$}_{\perp a})
\!\left[ \frac{d\tilde \sigma}{d\tilde t}(+\mbox{\boldmath $k$}_{\perp a}) -
\frac{d\tilde \sigma}{d\tilde t}(-\mbox{\boldmath $k$}_{\perp a}) \right]
\label{kint}
\eeq
\nd
where $d\tilde \sigma/d\tilde t$ means that now the partonic
cross section includes $\mbox{\boldmath $k$}_{\perp a}$ effects.

To give numerical estimates we need a model for the non perturbative
functions $I^{a/p}_{+-}(x,\mbox{\boldmath $k$}_{\perp})$; these non
diagonal distribution functions play for spin observables the same r\^ole
plaid by the usual diagonal distribution functions $G_{a/p}$ in unpolarized
cross-sections. We parameterize their $x$ dependence with simple power
behaviours. The dependence on $\mbox{\boldmath $k$}_{\perp}$ is treated, at
this stage, in a simplified way: we replace the integral in Eq.~(\ref{kint})
by the value of the integrand at some average $k_{\perp a} = \langle
\mbox{\boldmath $k$}^2_{\perp a} \rangle^{1/2}$. That is we set

\beq
\int \!d\mbox{\boldmath $k$}_{\perp a} \,
I^{a/p}_{+-}(x_a,\mbox{\boldmath $k$}_{\perp a}) \,
\frac{d\tilde \sigma}{d\tilde t}(\mbox{\boldmath $k$}_{\perp a}) =
\frac{\hat k_{\perp}}{M_h} N_a x_a^{\alpha_a} (1-x_a)^{\beta_a} \,
\frac{d\tilde \sigma}{d\tilde t}(k_{\perp a})
\label{param}
\eeq
\nd
where $M_h$ is some hadronic mass scale, of the order of 1 GeV, and
we assume $\hat k_{\perp} \simeq 0.5$ GeV/$c$. $N_a$ can be taken
from the usual distribution functions \cite{abm}.

Our final expression for the single spin asymmetry $A_N$ is then

\beq
A_N = \frac{
\sum_{abcd} \int dx_a dx_b
\frac{1}{x_c} I^{a/p}_{+-}(x_a,k_{\perp_a}) G^{b/p}(x_b)
\left[ \frac{d\tilde \sigma}{d\tilde t}(k_{\perp_a}) -
\frac{d\tilde \sigma}{d\tilde t}(-k_{\perp_a}) \right]
D_{\pi/c}(x_c)}
{2 \sum_{abcd} \int dx_a dx_b
\frac{1}{x_c} G^{a/p}(x_a) G^{b/p}(x_b)
\frac{d\hat\sigma}{d\hat t}(ab\to cd) D_{\pi/c}(x_c)}\cdot
\label{final}
\eeq

In Eq.~(\ref{final}) we take into account, at lowest perturbative QCD order,
all pos\-sible elementary interactions involving quarks and gluons. According
to $SU(6)$ proton wave functions we take $I^{u/p}_{+-}>0$ for $u$ quarks and
$I^{d/p}_{+-}<0$ for $d$ quarks (see footnote after Eq.~(\ref{iap})); the
sign of $I^{a/p}_{+-}$ for the other partonic contributions is less relevant,
and for the moment we assume all these contributions to be positive.
However, a more careful analysis is in progress \cite{abm}.
The unpolarized distribution and fragmentation functions are taken
from Ref. \cite{bro,field}.

In Fig.~1 we compare our results, at $p_T = 2$ GeV/$c$, with the experimental
data. Most contributions come from $qg \to qg$ and $gg \to gg$ processes and
the parameters $\alpha$ and $\beta$ of Eq.~(\ref{param}) yielding these
results are $\alpha_u = \alpha_d = \alpha_g \simeq -0.6$, $\beta_u = \beta_d
= 2.5$ and $\beta_g = 3.5$. We also find that, at $x_F=0$, $A_N \simeq 0$,
independently of $p_T$, in agreement with the most recent data \cite{nur}.

Our results clearly show how a careful treatment of spin observables and
the inclusion of intrinsic $k_\perp$ effects can yield sizeable values
of single spin asymmetries in hadronic inclusive pion production, via
perturbative QCD dynamics, contrary to widespread belief.

The off-diagonal distribution functions $I_{+-}^{a/p}$ introduced in
Eq.~(\ref{iap}) contain all the relevant non perturbative information;
similarly to the parton distribution functions in unpolarized processes,
they cannot be computed, but have to be taken from experiment. This is
essentially what we have done here, resulting in reasonable expressions
for $I_{+-}^{a/p}$; further discussions can be found in Ref. \cite{abm}.
Once the non diagonal distribution
functions have been obtained from one set of experiments, they can be
used to make genuine perturbative QCD predictions for other spin
observables like single spin asymmetries in $\pi+p^\uparrow \to \pi + X$ and
$p^\uparrow + p \to \gamma + X$. Some experimental results are already
available and more are soon expected.

\vskip24pt

\baselineskip=12pt

\normalsize

\nd {\Large \bf Figure Caption}

\begin{description}
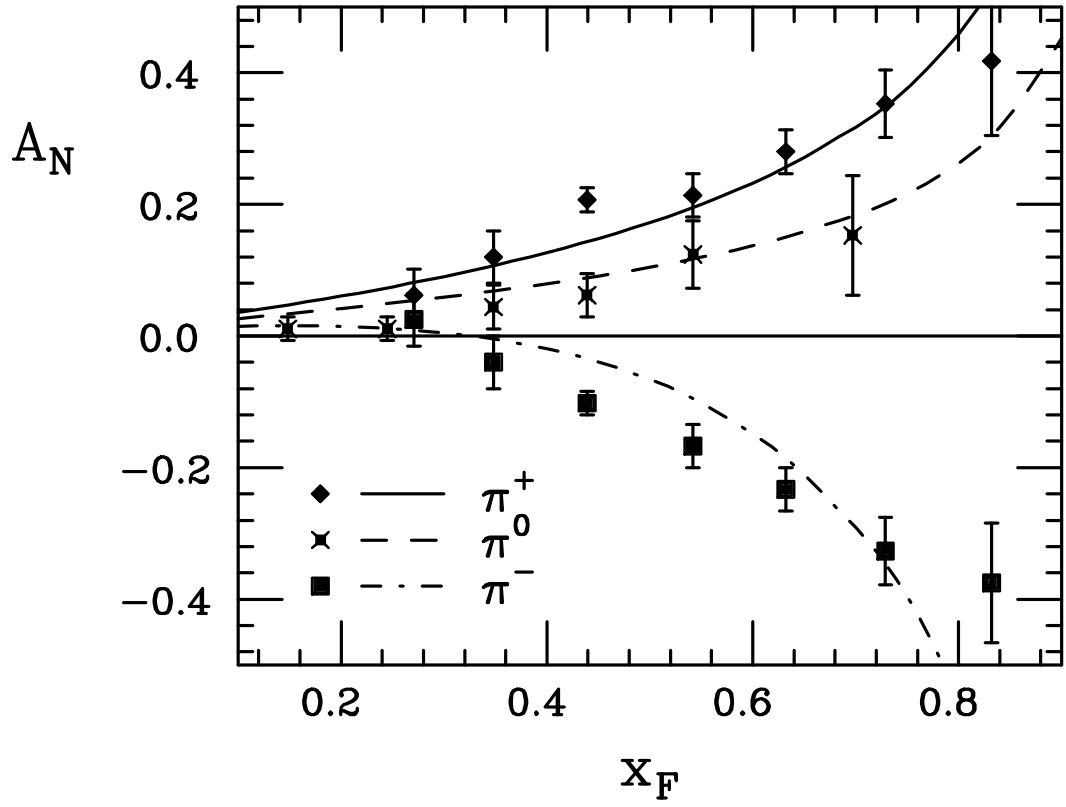
\item[Fig.~1] Single spin asymmetry for
$\pi ^+,\pi ^0,\pi ^-$, {\it vs.} $x_F$ at $p_T=2$ GeV/$c$,
from Eq.~(\protect\ref{final}),
compared to experimental results \protect\cite{e704a,e704b}
(see text for details).
\end{description}

\end{document}